\documentclass[aps,pra,twocolumn,showpacs]{revtex4}

\usepackage{dcolumn}
\usepackage{bm}
\usepackage{graphicx}
\usepackage{amsmath}
\usepackage{latexsym}
\usepackage{amsfonts}
\usepackage{amssymb}
\usepackage{array}

\newcommand{\ket}[1]{\left\vert#1\right\rangle}
\newcommand{\bra}[1]{\left\langle#1\right\vert}
\newcommand{\inner}[2]{\left\langle#1\kern-\nulldelimiterspace\left|#2\kern-\nulldelimiterspace\right.\right\rangle}

\begin{document}

\title{Generating a Schr\"odinger-cat-like state via a coherent superposition of photonic operations}

\author{Chang-Woo Lee$^{1,2,3}$, Jinhyoung Lee$^2$, Hyunchul Nha$^1$, and Hyunseok Jeong$^3$}
\affiliation{
$^1$Department of Physics, Texas A\&M University at Qatar, PO Box 23874, Doha, Qatar \\
$^2$Department of Physics, Hanyang University, Seoul 133-791, Korea \\
$^3$Center for Macroscopic Quantum Control \& Department of Physics and Astronomy, Seoul National University, Seoul, 151-747, Korea 
}

\date{\today}

\begin{abstract}
We propose an optical scheme to generate a superposition of coherent states with enhanced size adopting an interferometric setting at the single-photon level currently available in the laboratory.
Our scheme employs a nondegenerate optical parametric amplifier together with two beam splitters so that
the detection of single photons at the output conditionally implements the desired superposition of second-order photonic operations.
We analyze our proposed scheme by considering realistic on-off photodetectors with nonideal efficiency in heralding the success of conditional events.
A high-quality performance of our scheme is demonstrated in
view of various criteria such as quantum fidelity, mean output energy, and measure of quantum interference.
\end{abstract}

\maketitle

\setlength\arraycolsep{1pt}

\section{Introduction}

Since Schr\"odinger envisioned a \emph{Gedankenexperiment} to illustrate some peculiar feature of quantum mechanics when applied to macroscopic objects \cite{CatParadox}, the so-called Schr\"odinger cat state has long been of great interest from a fundamental point of view.
Furthermore, the Schr\"odinger cat state has also been identified as a useful resource for practical applications, e.g., in the field of quantum information science over the past decades.
There have been a great deal of theoretical and experimental attempts to generate a Schr\"odinger-cat-type state in various physical systems including
atomic system \cite{Monroe,Leibfried}, solid-state system \cite{Hofheinz}, and optical system \cite{Gao,Neergaard-Nielsen,Ourjoumtsev06,Ourjoumtsev07,Takahashi,Sasaki,Gerrits,Ourjoumtsev09}.
Among these, an optical-cat-like state,
especially a superposition of coherent states (SCS) \cite{Yurke86,Schleich}, turns out to be useful for
quantum computation \cite{Cochrane,JeongKim,Ralph03,Lund08,Marek10,Tipsmark},
quantum teleportation \cite{Enk,JKL,Wang}, 
quantum-enhanced metrology \cite{Gerry01,Ralph02,Munro02,Campos,Hirota,Joo}, and
fundamental test of quantum mechanics \cite{Gerry95,Munro00,Wilson,JSKAB,Stobinska,Gerry07,Lee09,Jeong09,Paternostro,Lee11}.
Along this line of practical significance,
a number of proposals have been made to generate an optical SCS \cite{Mecozzi,Yurke86,Yurke90,Dakna,Gerry99,Lund04,JKRH,Jeong05,Lance,Glancy,Molmer,Marek08,He}, which subsequently led to considerable experimental progresses \cite{Neergaard-Nielsen,Ourjoumtsev06,Ourjoumtsev07,Takahashi,Sasaki,Gerrits,Ourjoumtsev09}.

The original proposals for generating an optical SCS in a deterministic way were based on Kerr nonlinear medium \cite{Mecozzi,Yurke86}.
However, their realizations have not yet been made since the high nonlinearity required for such a deterministic scheme is not achievable with the current technology.
Alternatively, most of the SCSs created so far in laboratory \cite{Neergaard-Nielsen,Ourjoumtsev06,Ourjoumtsev07,Ourjoumtsev09,Takahashi,Sasaki,Gerrits} are based on a nondeterministic scheme that generates a target state only under prespecified conditions. For instance, the photon-subtraction method \cite{Dakna} can implement a SCS
relying on the fact that
even (odd)-photon-subtracted squeezed vacuum states well approximate even (odd) SCSs with small amplitudes.
Marek {\it et al.} also investigated the effect of photon addition as well as subtraction on the initial squeezed vacuum state
to obtain an output state close to a squeezed SCS \cite{Marek08}.
On the other hand, the experimental scheme of Ref. \cite{Ourjoumtsev07}
produced a squeezed SCS by conditioning an input Fock state on the outcome of homodyne detection. That is, if a
Fock state $\ket{n}$ is injected into a beam splitter and conditioned on a certain predetermined value of the quadrature amplitude at one output mode, the  other output state converges well to a 3-dB-squeezed SCS as $n$ becomes large.

In this paper we propose an experimental scheme to generate a squeezed SCSs with enhanced size employing a coherent superposition of photonic operations.
Our proposal makes use of an interferometic setting that combines a nondegenerate parametric amplifier (NDPA) together with two beam splitters.
The detection of single photons at two output modes conditionally implements a coherent superposition of second-order operations, namely, $\hat{a}\hat{a}^\dag$ and $\hat{a}^{\dag2}$, on an arbitrary input state. 
We show that this scheme, when applied to a single-photon input, can successfully generate an output state very close to an odd-parity squeezed SCS which includes a triple-photon component or higher.
To illustrate the feasibility of our scheme, we consider the use of nonideal on-off photodetectors for heralding the conditional events and demonstrate a high-quality performance in terms of various criteria such as quantum fidelity, mean output energy, and the measure of quantum intereference \cite{MQI}.
Our proposed scheme thus seems very feasible within the currently available experimental techniques \cite{Kim08}.
In particular, we note that a single-photon interferometric scheme similar to ours was recently proposed \cite{PhotonAdd} and also experimentally realized \cite{CommRel} to verify the bosonic commutation relation $[\hat{a},\hat{a}^\dag]=1$.

This paper is organized as follows.
In Sec. \ref{squeezedSCS}, we briefly introduce a class of squeezed SCSs targeted in our work.
In Sec. \ref{scheme}, we propose an experimental scheme to implement a superposition operation to our end.
With a brief account of the working principle under our scheme, its rigorous description is also provided within the Wigner-function formalism.
In Sec. \ref{performance}, we evaluate the performance of our scheme in terms of various criteria such as quantum fidelity, mean output energy, and measure of quantum interference. In Sec. \ref{inefficiency}, our analysis is further extended to investigate the effect of on-off photo detectors with nonideal efficiency.
Finally, our results are summarized in Sec. \ref{remark}.

\section{Squeezed SCS}
\label{squeezedSCS}

Although there exist some generalized versions of SCSs \cite{Yurke86,Schleich},
we begin with two typical SCSs, namely, \emph{even} and \emph{odd} SCSs,
\begin{equation}
\ket{\mathrm{SCS}_{\pm} (\alpha)} = \mathcal{N}_{\pm} \left( \ket{\alpha} \pm \ket{-\alpha} \right),
\label{SCS}
\end{equation}
where
$
\mathcal{N}_{\pm} = 1/ \sqrt{2 \pm 2e^{-2 \alpha^2}}
$
are the normalization factors and $\ket{\pm \alpha}$ is a coherent state of amplitude $\pm \alpha$.
Here $\alpha$ is assumed to be real for simplicity but without loss of generality.
The SCS with the positive (negative) sign in Eq. (\ref{SCS}) is called
an even (odd) SCS because it contains only even (odd)-numbered Fock states for any $\alpha$.
The size of SCS may be quantified by the magnitude of the amplitude $|\alpha|$, which also characterizes the distance between the two component states
$\ket{\alpha}$ and $\ket{-\alpha}$ in phase space.

Suppose now that we apply a squeezing operation $\hat{S} \left( s \right) = \exp \left[ \frac{s}{2} \left( \hat{a}^2 - \hat{a}^{\dag 2} \right) \right]$
($s$: real) on the SCS in Eq. (\ref{SCS}).
Note that $\hat{S} \left( s \right)$ squeezes a quantum state along the real (imaginary) axis in phase space for a positive (negative) $s$.
When applied to $\ket{\mathrm{SCS}_{\pm}}$ in Eq. (\ref{SCS}),
$\hat{S} \left( s \right)$ does not change the parity of
$\ket{\mathrm{SCS}_{\pm}}$ since $\hat{S} \left( s \right)$ is invariant under the parity operator
$(-1)^{\hat{a}^{\dag} \hat{a}}$. Hereafter, we label the state
\begin{equation}
\ket{s \rm SCS_{\pm}(\alpha)} = \hat{S} (s) \ket{\rm SCS_{\pm}(\alpha)}
\label{sSCS}
\end{equation}
as a {\it squeezed} (even or odd) SCS ($s$SCS).
In view of Eq. (\ref{sSCS}), let us now define a class of states, 
\begin{equation}
\ket{\psi_n} = \hat{S}  (\ln \sqrt{2})\ket{{\rm SCS}_{(-)^n} (\sqrt{n})},
\label{psi}
\end{equation}
which is 3-dB-squeezed SCS and has the parity $(-1)^n$ and amplitude $\alpha = \sqrt{n}$ for a non-negative integer $n$.

Ourjoumtsev {\it et al.} proposed a novel conditional scheme to create
a similar Shr\"{o}dinger-cat-type state, which generated a state close to $\ket{\psi_2}$ in Eq.~(\ref{psi}).
Their idea was to utilize homodyne-conditioning technique
with $n$-photon state $\ket{n}$ as an input \cite{Ourjoumtsev07}:
The state $\ket{n}$ is injected into a 50:50 beamsplitter with the other input in a vacuum state.
The output state is kept conditioned on a particular outcome of homodyne detection at the other mode, which produces a class of states $\ket{\phi_n}$.
Along with the trivial states
$\ket{\phi_0} = \ket{0}$ and $\ket{\phi_1} = \ket{1}$,
other lowest states turn out to be
\begin{eqnarray}
\ket{\phi_2} &=& \sqrt{\frac{1}{3}} \ket{0} + \sqrt{\frac{2}{3}} \ket{2},
\label{phi_2} \label{phi2} \\
\ket{\phi_3} &=& \sqrt{\frac{3}{5}} \ket{1} + \sqrt{\frac{2}{5}} \ket{3}, \label{phi_3}\\
\ket{\phi_4} &=& \sqrt{\frac{3}{35}} \ket{0} + \sqrt{\frac{24}{35}} \ket{2} + \sqrt{\frac{8}{35}} \ket{4},
....
\end{eqnarray}
It is worth noting that
$\ket{\phi_n}$ has the same parity as the Fock state $\ket{n}$:
If $n$ is even (odd), $\ket{\phi_n}$ has only even (odd) number of photons just like $\ket{\mathrm{SCS}_{\pm}}$ and $\ket{s \mathrm{SCS}_{\pm}}$.

Interestingly,
$\ket{\phi_n}$ is very close to $\ket{\psi_n}$ in view of quantum fidelity:
$
F \equiv \left| \inner{\psi_n}{\phi_n} \right|^2 \ge 0.97
$
for all $n\ge0$, and $F \to 1$ as $n \to \infty$.
Ourjoumtsev {\it et al.} produced a state targeting $\ket{\phi_2}$ \cite{mistake}
with a two-photon state $\ket{2}$ as an input. The target state $\ket{\phi_2}$ is very close to $\ket{\psi_2}$ with the fidelity $F=$ 97.2\% and
can be made even closer to $\ket{s \rm SCS_{\pm}}$ with $F=$ 99.0\% by increasing the squeezing level from 3 to 3.5 dB.
In the next sections, we propose an experimental scheme to generate a larger-size $s$SCS, i.e., $\ket{\phi_3}$ and beyond.

\section{Scheme to generate SCS}
\label{scheme}
In this section, we present an experimental scheme to implement a coherent superposition of photonic operations for generating a $s$SCS.
For instance, to generate the target state $\ket{\phi_3}$, which is a superposition of the number states $\ket{1}$ and $\ket{3}$,
one has to apply a superposition of identity operation $I$ and two-photon addition $\hat a ^{\dag 2}$ to an input single-photon state $|1\rangle$.
That is, $(tI+r\hat a ^{\dag 2})|1\rangle\sim c_1|1\rangle+c_2|3\rangle$.
Our scheme makes use of a NDPA together with two beam splitters, and the output state is kept on the condition of single-photon detections at two output modes of a beam splitter (Fig. 2). This implements the superposition operation of the form $t\hat a\hat a ^{\dag}+r\hat a ^{\dag 2}$, as will be shown below. Note that the effect of the operation $\hat a\hat a ^{\dag}$ is essentially the same as that of the identity operation $I$ for the case of Fock-state input.

To begin with, we describe our experimental scheme under the small-parameter approximations \cite{PhotonAdd,CommRel,LeeNha},
which characterizes the output state when using an NDPA with small squeezing.
In Sec. III B, we will more rigorously describe our proposed scheme using a Wigner-function formalism in phase space.

\subsection{Brief sketch of the scheme}
\label{Sketch}

\begin{figure}[t]
\centerline{\includegraphics[width=\columnwidth]{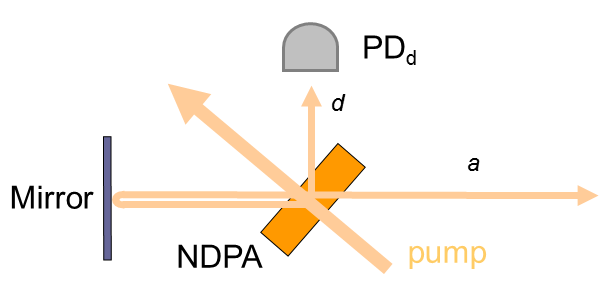}}
\caption{(Color online)
Experimental scheme for a preliminary single-photon generation.
The pump beam creates twin beams as it passes through a NDPA.
One output mode is directed to the detector PD$_d$ and the other to a mirror.
A detection event at PD$_d$ heralds the generation of an approximate single-photon state in mode $a$.
}
\label{scheme1}
\end{figure}

The first stage of our scheme is to prepare a single-photon state $\ket{1}$.
This can be done by pumping an NDPA with both inputs of signal and idler modes in vacuum states and then by projecting the output idler to a single-photon state (Fig. \ref{scheme1}). More precisely, the interaction between the signal mode $a$ and the idler mode $d$ within the NDPA is described
by a two-mode squeezing operator
\begin{equation}
\hat{S}_{ad} (s) = \exp [s ( \hat{a}  \hat{d} -  \hat{a}^\dagger \hat{d}^\dagger )].
\end{equation}
With the two inputs in vacuum states, the output becomes
\begin{eqnarray}
\ket{\Psi}_{ad}& =& \hat{S}_{ad} \left( s \right) \ket{0}_a \ket{0}_d\nonumber\\
&\approx&
\left( 1 - s \hat{a}^\dagger  \hat{d}^\dagger   \right) \ket{0}_a \ket{0}_d  \\
&=& \ket{0}_a \ket{0}_d  - s \ket{1}_a \ket{1}_d ,
\end{eqnarray}
where the condition $s\ll1$ (small squeezing) is used.
Now, the detection of a single photon at mode $d$ gives the desired single-photon state in mode $a$:
\begin{equation}
\ket{\Psi}_{a} \sim \bra{1}_d \left( \ket{\Psi}_{ad} \right) \sim \ket{1}_a .
\label{single-photon}
\end{equation}

\begin{figure}[t]
\centerline{\includegraphics[width=\columnwidth]{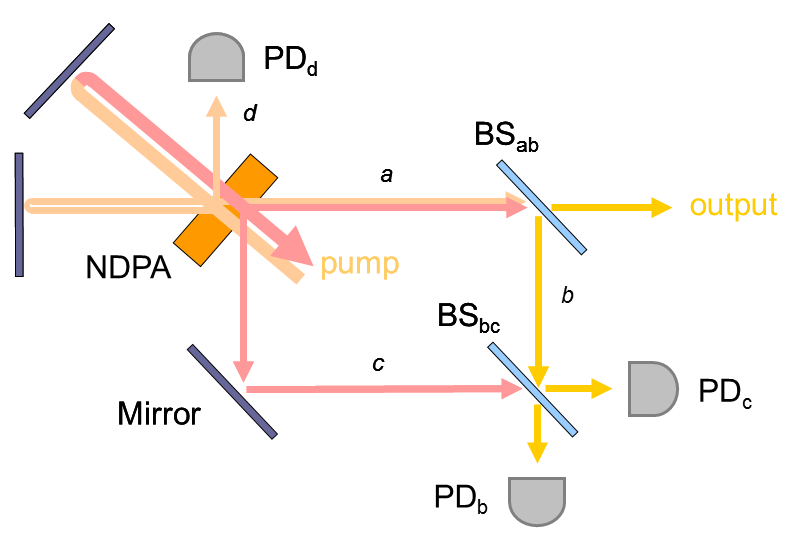}}
\caption{(Color online)
A whole process to generate a desired state $\ket{\phi_3}$ and beyond.
After generating the approximate single-photon state,
the pump beam is redirected by a mirror into the NDPA and creates another twin beams.
The coincident detection of single photons at PD$_b$ and PD$_c$ heralds a successful implementation of a superposition operation coherently combining
$\hat{a}\hat{a}^\dag$ and $\hat{a}^{\dag2}$. See the main text.}
\label{scheme2}
\end{figure}
In Fig. \ref{scheme1}, the signal mode $a$ is reflected by a mirror to be injected again into the same NDPA (see Fig. 2).
By doing this, we can reduce the number of required NDPAs, and furthermore, as the input signal has been produced within the same NDPA,
the mode-matching condition, which is critical for quantum interference, may be better satisfied.
Thus, in the next stage, we reuse the pump beam during its coherence time for the NDPA process.
Under the small-squeezing condition, the output now reads
\begin{eqnarray}
\ket{\Psi}_{ac} &&=
\hat{S}_{ac} \left( s \right) \ket{\Psi}_{a} \ket{0}_c  \\
\approx&&
\left( 1 - s \hat{a}^\dagger  \hat{c}^\dagger + \frac{1}{2} s^2 \hat{a}^{\dagger 2} \hat{c}^{\dagger 2}  \right)
\ket{1}_a  \ket{0}_c
\label{Psi_ac_approx}
\\ =&&
\ket{1}_a  \ket{0}_c -
\sqrt{2} s \ket{2}_a  \ket{1}_c + \sqrt{3} s^2 \ket{3}_a  \ket{2}_c .
\label{Psi_ac_result}
\end{eqnarray}
Note that other irrelevant terms are not included in Eq. (\ref{Psi_ac_approx}) and that
only the last two terms in Eq. (\ref{Psi_ac_result}) will survive the subsequent detection process as described below.

In the third stage,
we tap a very small amount of the beam in mode $a$ using a beamsplitter BS$_{ab}$ whose transmittance is very close to 1.
This can be described by a beamsplitter operator $\hat{B}_{ab} (t)$ acting on the state $\ket{\Psi}_{ac}$, with mode $b$ in a vacuum state, as
\begin{equation}
\ket{\Psi}_{abc} = \hat{B}_{ab} ( t_1 ) \ket{\Psi}_{ac} \ket{0}_b,
\label{Psi_abc}
\end{equation}
where
\begin{equation}
\hat{B}_{ab} (t) = \exp \left[ \frac{\theta}{2} \left( \hat{a} \hat{b}^\dagger - \hat{a}^\dagger \hat{b} \right) \right]
\end{equation}
with its transmissivity (reflectivity) given by
$t = \cos \frac{\theta}{2}$ $\left(r = \sin \frac{\theta}{2}\right)$.
Under the condition $r_1\ll1$ ($t_1\approx1$) and $O(r_1)\sim O(s)$, Eq. (\ref{Psi_abc}) may be approximated as
\begin{eqnarray}
\ket{\Psi}_{abc} \approx&&
\left( 1+ r_1 \hat{a} \hat{b}^\dagger \right) \ket{\Psi}_{ac} \ket{0}_b \\
\sim &&
-2 s r_1 \ket{1}_a \ket{1}_b \ket{1}_c + \sqrt{3} s^2 \ket{3}_a \ket{0}_b \ket{2}_c .
\label{Psi_abc_approx}
\end{eqnarray}

\begin{figure}[t]
\centerline{\includegraphics[width=\columnwidth]{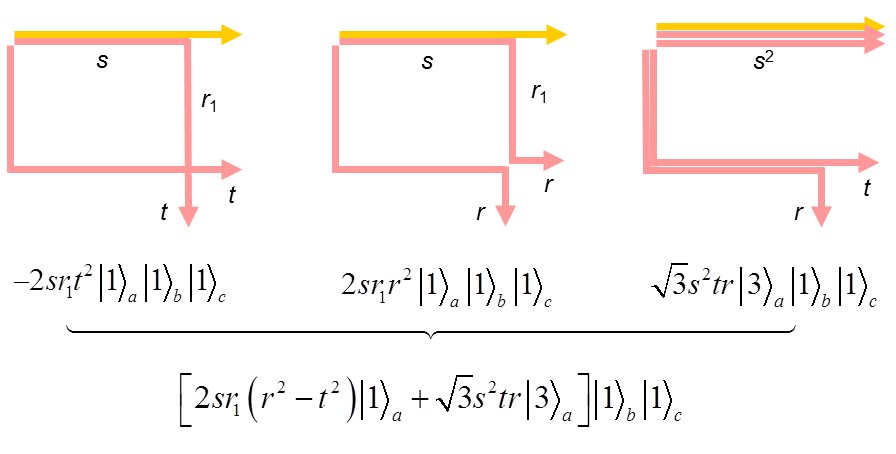}}
\caption{(Color online)
The first two and the third diagrams show a single- and triple-photon generation, respectively.
Yellow arrows represent the single-photon generated in the first stage of our scheme in Fig. 2, whereas
red arrows represent a single-photon in the first and the second diagram, or a two-photon in the third diagram, generated in each mode $b$ and $c$.
The parameter $s$ denotes the degree of squeezing within the NDPA, $r_1$ the reflectivity of the beamsplitter BS$_{ab}$,
and $t$ ($r$) the transmissivity (reflectivity) of the beamsplitter BS$_{bc}$.
In the first two diagrams,
$s$ indicates that twin single-photons are generated and
$r_1$ indicates that upper one of twin beams is reflected by BS$_{ab}$.
Two $t$'s ($r$'s) in the first (second) diagram means that
two beams all pass through (are reflected at) BS$_{bc}$ without reflection (transmission).
In the third diagram,
$s^2$ indicates that twin two-photons are generated and
$t$ and $r$ means that one of the two photons in the lower beam passes through, and
the other is reflected, at BS$_{bc}$.
}
\label{scheme3}
\end{figure}

Finally, after the state $\ket{\Psi}_{abc}$ passes through another beam-splitter BS$_{bc}$ (transmissivity $t$),
we obtain the output state
\begin{eqnarray}
\ket{\Psi}_{abc} ' &=& B_{bc} ( t ) \ket{\Psi}_{abc}\\
&\approx&
2 s r_1 ( r^2 -t^2 )\ket{1}_a \ket{1}_b \ket{1}_c\nonumber \\&&+
\sqrt{3} s^2 t r \ket{3}_a \ket{1}_b \ket{1}_c.
\label{Psi'_abc}
\end{eqnarray}
Note that
the first term in (\ref{Psi'_abc}) is due to the first term in (\ref{Psi_abc_approx}):
The factor $t^2$ ($r^2$) indicates that
two single photons are simultaneously transmitted (reflected) via the BS$_{bc}$, and the different sign between $r^2$ and $t^2$ arises due to the phase change induced by two reflections at the beamsplitter.
The last term in (\ref{Psi'_abc}) implies that
two photons of mode $c$ in (\ref{Psi_abc_approx}) are split into individual single photons in mode $b$ and $c$.
The whole process here is a quantum-mechanical superposition of three different paths,
which is illustrated by the diagrams in Fig. \ref{scheme3}.

Now, if both detectors ${\rm PD}_b$ and ${\rm PD}_c$ simultaneously register single photons,
the conditional output is given by
\begin{eqnarray}
\ket{\Psi}_{\rm out}  &\sim&
\bra{1}_b \bra{1}_c \left( \ket{\Psi}_{abc} '  \right) \nonumber \\
&\approx& 2 s r_1 ( r^2 -t^2 )\ket{1}_a +
\sqrt{3} s^2 t r \ket{3}_a .
\label{Output}
\end{eqnarray}
Thus, by adjusting the parameters $s,~ t_1(r_1)$, and $t (r)$ in the experimental scheme of Fig. 2,
we can obtain $\ket{\phi_3}$ in Eq. (\ref{phi_3}).

More generally, if an arbitrary input state $\ket{\Psi}_{\rm in}$ is injected into the scheme in Fig. 2,
one can similarly confirm that the output state is given by
\begin{eqnarray}
\ket{\Psi}_{\rm out}&=&\bra{1}_b \bra{1}_c \hat{B}_{bc} (t)\hat{B}_{ab} ( t_1 ) \hat{S}_{ac} \ket{\Psi}_{\rm in}\ket{0}_{b}\ket{0}_{c} \nonumber\\
 &\approx&
 \left( C_1\hat{a}\hat{a}^\dag  +C_2
 \hat{a}^{\dag 2} \right) \ket{\Psi}_{\rm in}
\end{eqnarray}
where $C_1=s r_1 ( r^2 -t^2 )$ and $C_2=\frac{1}{\sqrt{2}} s^2 t r$.
Therefore, our scheme in Fig. 2 implements a superposition of second-order operations $\hat{a}\hat{a}^\dag$ and $\hat{a}^{\dag 2}$ on an arbitrary input.
In principle, one can generate $\ket{\phi_3},~\ket{\phi_5},~\ket{\phi_7},...$ or $\ket{\phi_2},~\ket{\phi_4},~\ket{\phi_6},...$
by consecutively applying this superposition operation to an input state $\ket{1}$ or $\ket{0}$, respectively.

\subsection{Rigorous formulation using Wigner-function approach}
\label{WignerFormalism}

In Sec. \ref{Sketch}, 
we have briefly described the working principle of generating the target $s$SCS under small-parameter approximations.
We now present an exact formulation of our experimental scheme in Fig. 2 using the Wigner-function formalism in phase space.
We here give a concise description of the Wigner-function method with essential steps and all detailed derivations are shown in Appendix B.

Let us start with the twin-beam state created within the NDPA, which is formally described by the two-mode squeezing of the input vacuum states $\ket{0}_a \ket{0}_d$. In phase-space representation,
a two-mode squeezing has the effect of pseudorotation on the arguments of two quasi-distribution functions. That is,
the input two-mode Wigner function
$W_{\rm vac} (\beta_a) W_{\rm vac} (\beta_d)$
transforms into
\begin{eqnarray}
W_{ad} \left( \beta_a ,\beta_d  \right) &=&
W_{\rm vac} \left(   \beta_a \cosh s +  \beta^*_d  \sinh s \right)  \nonumber \\
&& \times \;
W_{\rm vac} \left(  \beta_d \cosh s + \beta^*_a  \sinh s  \right) \nonumber \\
&=& (2/ \pi)^2\exp \big[ - 2\cosh 2s\big( \left| \beta_a  \right|^2  + \left| \beta_d  \right|^2  \big)  \nonumber \\
& &+ 2\sinh 2s\left( \beta_a \beta_d  + \beta^*_a  \beta^*_d  \right)  \big] ,
\label{W_ad}
\end{eqnarray}
where
$
W_{\rm{vac}}  \left( \beta   \right)  = (2/ \pi) \exp(- 2 |\beta|^2)
$
is the Wigner function of a vacuum state $\ket{0}$, with $\beta = \beta_r + i \beta_i$.

If the detector ${\rm PD}_d$ in Fig. 2 registers any number of photons, the conditional output state is given by
$\rho_a\sim {\rm Tr}_{d}\left[(I-\ket{0}\!\bra{0})_d\,\rho_{ad}\right]$, whose Wigner function becomes
\begin{eqnarray}
W_{a} \left( \beta_a   \right) &=
( 1/P_d ) \int {d^2 \beta_d }
W_{ad} \left( {\beta_a ,\beta_d } \right)  \nonumber \\
& \times \;
\left[ {1 - \pi  W_{{\rm{vac}}} \left( \beta_d  \right)} \right],
\label{W_a}
\end{eqnarray}
with its normalization
\begin{eqnarray}
P_d&=& {\rm Tr}_{ad}[I_a \otimes (I-\ket{0}\!\bra{0})_d \,\rho_{ad}]\nonumber\\
&=& \int \! {d^2 \beta_a } \int \! {d^2 \beta_d } \,
W_{ad} \left( {\beta_a ,\beta_d } \right)  \left[ {1 - \pi  W_{{\rm{vac}}} \left( \beta_d  \right)} \right].
\label{P_d}
\end{eqnarray}
Note that $P_d$ represents the conditional probability for detection events and its detailed funtional form is given in Appendix: B.
Equation (\ref{W_a}) is an exact version of Eq. (\ref{single-photon}).

The conditional state in (\ref{W_a}) is again injected to the NDPA with other mode $c$ in a vacuum state. We thus obtain an output state from the NDPA,
whose Wigner function is given by
\begin{eqnarray}
W_{ac} \left( \beta_a ,\beta_c  \right) &=&
W_{a} \left(   \beta_a \cosh s +  \beta^*_c  \sinh s \right)  \nonumber \\
&& \times \;
W_{\rm{vac}} \left(  \beta_c \cosh s + \beta^*_a  \sinh s  \right).
\label{Wac}
\end{eqnarray}

The third step, i.e.,
mixing the mode $a$ with another mode $b$ (vacuum state) at a beamsplitter $B_{ab}$($t_1^2 \approx 1$ or $r_1^2 \ll 1$) is described by
\begin{eqnarray}
W_{abc} \left( \beta_a , \beta_b , \beta_c  \right) &=&
W_{ac} \left( t_1 \beta_a  + r_1 \beta_b , \beta_c \right)  \nonumber \\
&& \times \;
W_{\rm{vac}} \left( t_1 \beta_b  - r_1 \beta_a  \right).
\label{1stBS}
\end{eqnarray}
Another mixing process at the beamsplitter $B_{bc}$ gives
\begin{equation}
W_{abc} ' \left( {\beta_a , \beta_b , \beta_c } \right) =
W_{abc} \left( \beta_a ,  t \beta_b  + r  \beta_c, t \beta_c  - r \beta_b \right) .
\label{2ndBS}
\end{equation}
Finally, the success probability for the coincident detection of photons at ${\rm PD}_b$ and ${\rm PD}_c$ is given, similar to Eq. (\ref{P_d}), by
\begin{eqnarray}
P_{bc} =
\int {d^2 \beta_a } \int  d^2 \beta_b \int d^2 \beta_c \,
W_{abc} ' \left( {\beta_a ,\beta_b ,\beta_c } \right)  \nonumber \\
\times
\left[ {1 - \pi W_{\rm{vac}} \left( \beta_b  \right)} \right]
\left[ {1 - \pi W_{\rm{vac}} \left( \beta_c  \right)} \right] .
\end{eqnarray}
In addition, the Wigner function of the conditional output is given by
\begin{eqnarray}
W_{\rm out} \left( \beta_a  \right) &= &
( 1/P_{bc} ) \int \! d^2 \beta_b \! \int d^2 \beta_c \,
W_{abc} ' \left( {\beta_a ,\beta_b ,\beta_c } \right)  \nonumber \\
& \times &
\left[ {1 - \pi W_{\rm{vac}} \left( \beta_b  \right)} \right]
\left[ {1 - \pi W_{\rm{vac}} \left( \beta_c  \right)} \right] .
\end{eqnarray}

\begin{figure}[t]
\centerline{ \bf{(a)} \hskip2.3cm \bf{(b)} \hskip2.3cm \bf{(c)}}
\centerline{\includegraphics[width=\columnwidth]{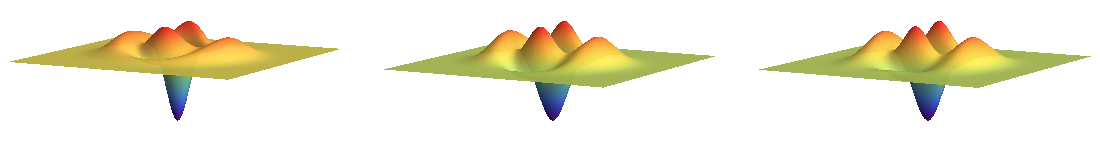}}
\caption{(Color online)
Wigner functions of (a) $\ket{\phi_3}$, (b) a state generated via the scheme in Fig. 2, and (c) $s$SCS [Eq.(\ref{sSCS})].
The state in (b) is produced with the parameters $s=0.04$ and $t_1^2=0.999$ ($r_1^2=0.001$).
The $s$SCS in (c) is 2.9-dB squeezed ($s'=0.33$) with $\alpha=1.7$.
The fidelity of the output state in (b) is 99\% with respect to the $s$SCS.
}
\label{WignerFunc}
\end{figure}

One of the typical examples generated under our scheme is shown in Fig. \ref{WignerFunc},
together with the target state $\ket{\phi_3}$ and $s$SCS [Eq.(\ref{sSCS})].
The figures show that the output state is slightly closer to $s$SCS than to $\ket{\phi_3}$.
We investigate the quantum fidelity with respect to $s$SCS as
\begin{equation}
F = \pi \int {d^2 \beta } \;W_{\rm out} \left( \beta  \right)
W_{s \rm SCS} \left( \beta  \right) ,
\label{fidelity}
\end{equation}
where
$W_{s \rm SCS} \left( \beta  \right)$
is the Wigner function of $s$SCS [Eq.(\ref{sSCS})] with squeezing parameter $s'$:
\begin{eqnarray}
W_{\rm SCS} (\beta) &=&
{\cal N}^{\,2}_{-} \big[
W_{\rm vac} (\beta_r - \alpha, \beta_i) + W_{\rm vac} (\beta_r + \alpha, \beta_i) \nonumber\\
&&-2\, W_{\rm vac} (\beta) \cos (4 \alpha \beta_i) \big] .
\label{WSCS} \\
W_{s \rm SCS} (\beta) &=&
W_{\rm SCS} \left( \beta \cosh s' + \beta^\ast \sinh s' \right).
\label{WsSCS}
\end{eqnarray}
In Fig. 4, the quantum fidelity of the output state is 99\% with respect to the $s$SCS.

\section{Performance of the scheme}
\label{performance}

\begin{figure}[t]
\centerline{ \bf{(a)} }
\vskip0.1cm
\centerline{\includegraphics[width=\columnwidth]{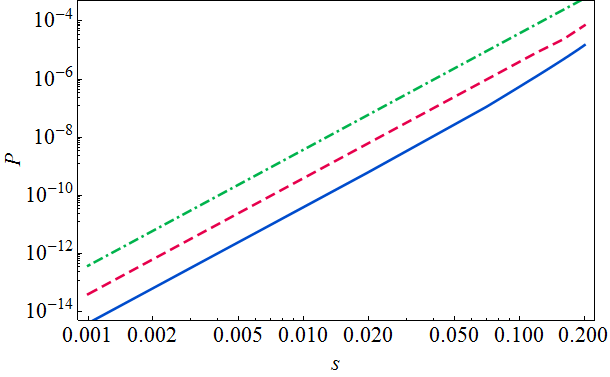}}
\vskip0.1cm
\centerline{ \bf{(b)} }
\vskip0.1cm
\centerline{\includegraphics[width=\columnwidth]{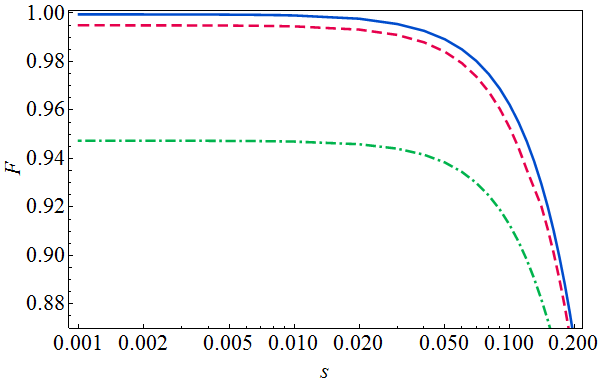}}
\caption{(Color online)
(a) success probability and (b) quantum fidelity of an output state as functions of the squeezing parameter $s$.
The three curves in each panel represent the cases of
$r_1^2 =$ 0.001 (blue solid), 0.01 (red dashed), and 0.1 (green dot-dashed), respectively.
The success probability $P$ is the joint detection probability of all three detectors in the scheme of Fig. \ref{scheme2}.
The fidelity $F$ quantifies the closeness between the generated state and the target $s$SCS, which is optimized over
$t$ ($r$) in (\ref{2ndBS}), $s'$ in (\ref{WsSCS}) and $\alpha$ in (\ref{WSCS}).
}
\label{P&F}
\end{figure}

In this section, we investigate the overall performance of our scheme in terms of various criteria.
In particular, we look into success probability, quantum fidelity, mean photon number, and a measure of quantum interference
to properly address the quality of an output state under our scheme.

We first investigate the success probability $P$ and
the fidelity $F$ in Eq. (\ref{fidelity}) in more detail.
The overall success probability $P$ is given by $P = P_d P_{bc}$,
which is the joint probability of generating a single-photon state in the first stage and
detecting photons simultaneously at PD$_b$ and PD$_c$ in the final stage.
Figure \ref{P&F} shows the success probability and the fidelity
as functions of the squeezing parameter $s$.
In each panel, three curves refer to the cases with different reflectances of the beams plitter ${\rm BS}_{ab}$ in Fig. \ref{scheme2}, i.e.,
$r_1^2 = 1-t_1^2 =$ 0.001 (blue solid), 0.01 (red dashed), and 0.1 (green dot-dashed).
Note that
the fidelity is optimized over $s'$, $\alpha$, and $t$ (or $r$) at fixed $s$ and $t_1$ and that
the success probability is evaluated corresponding to those parameters.

Some general trends can be seen from Figs. \ref{P&F}(a) and \ref{P&F}(b).
First, the success probability increases with the squeezing parameter $s$ whereas the quantum fidelity decreases with $s$.
Second, similarly, increasing the reflectance $r_1^2$ enhances the success probability but lowers the fidelity as a trade-off.
A higher squeezing produces more twin-beam photons within the NDPA,
hence more chances of detecting photons at the detectors PD$_{b}$ and PD$_{c}$.
Similar reasoning can be given to the trend with the reflectance of BS$_{ab}$.
On the other hand, the degradation of quantum fidelity with $s$ and $r_1$ is somewhat expected as our scheme is envisioned under the condition
that the squeezing $s$ is small and that $r_1$ should also be small in the same order of $s$,
as discussed in Sec. \ref{Sketch}.

Even though the quantum fidelity can be very high, it would be practically not useful if the conditional probability is too low.
However, in our scheme,
if the fidelity above 90\% is acceptable, the success probability goes up to $10^{-6}-10^{-5}$ for $r_1^2 =$ 0.001 or 0.01.
Thus, if the repetition rate of a pulsed beam is assumed to be around 1 MHz,
the generation rate becomes 1-10 Hz.
Therefore, with moderate values of $s$ and $r_1$,
the present scheme seems quite feasible for an experimental realization of high-fidelity $s$SCS.

\begin{figure}[t]
\centerline{ \bf{(a)} }
\vskip0.1cm
\centerline{\includegraphics[width=\columnwidth]{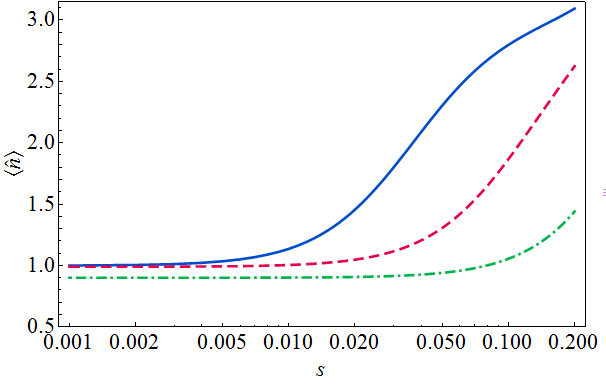}}
\vskip0.1cm
\centerline{ \bf{(b)} }
\vskip0.1cm
\centerline{\includegraphics[width=\columnwidth]{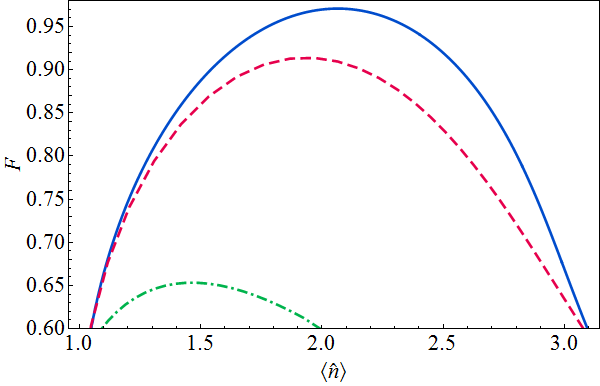}}
\caption{(Color online)
(a) Mean photon number $\langle \hat{n} \rangle$
as a function of the degree of squeezing $s$ for the cases of
$r_1^2 =$ 0.001 (blue solid), 0.01 (red dashed), and 0.1 (green dot-dashed), respectively.
Here the parameters used for optimizing the fidelity in Fig. \ref{P&F} are fixed as
$t=0.21$, $s'=0.57$, and $\alpha=2.4$. 
Note that $\langle \hat{n} \rangle$ increases with $s$ but decreases with the reflectance $r_1^2$.
Since a larger squeezing degrades the fidelity [see Fig. \ref{P&F}(b)],
the fidelity as a function of $\langle \hat{n} \rangle$ is also plotted in (b).
With a fixed target state, there is a critical value of $s$ which gives the maximum value for the fidelity.
Thereafter, $\langle \hat{n} \rangle$ increases, but the fidelity decreases, with $s$.
}
\label{MeanNumber}
\end{figure}

Next, as a size criterion,
we investigate the mean photon number $\langle \hat{n} \rangle$ of the generated state.
As can be seen from Fig. \ref{MeanNumber}(a),
a larger squeezing $s$ and a smaller reflectance $r_1^2$ give a larger size of the generated state,
which may be expected from the relative ratio of the component states in Eq. (\ref{Output}).
Note that the mean photon number $\langle \hat{n} \rangle$ can go over 3,
which is relatively large compared to 2.75 of the recently generated state \cite{Gerrits}.
However, as can be seen from Fig. \ref{MeanNumber} (b),
in order to obtain a larger size state,
the fidelity should be sacrificed to some degree.
Nevertheless, with the size being equal to 2.75,
the fidelity 83\% under our scheme is still higher than 59\% in Ref. \cite{Gerrits}.

\begin{figure}[t]
\centerline{ \bf{(a)} }
\vskip0.1cm
\centerline{\includegraphics[width=\columnwidth]{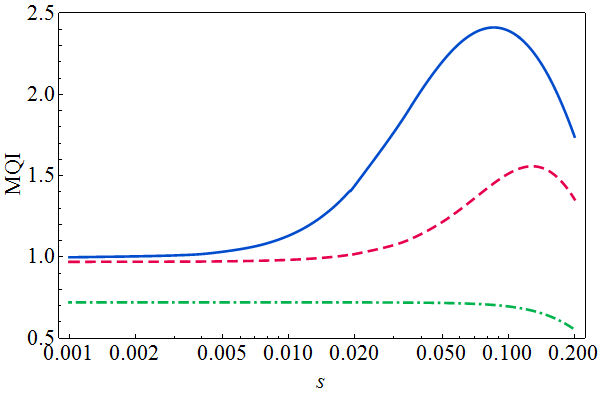}}
\vskip0.1cm
\centerline{ \bf{(b)} }
\vskip0.1cm
\centerline{\includegraphics[width=\columnwidth]{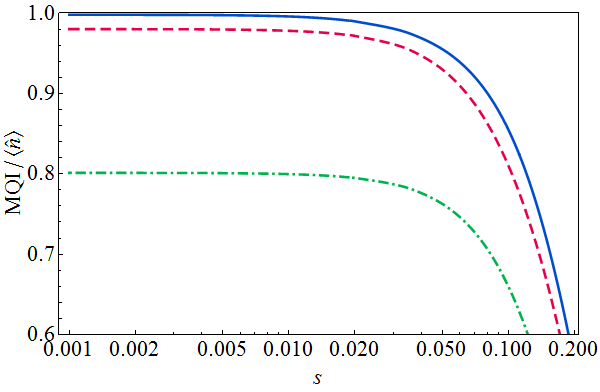}}
\caption{(Color online)
(a) MQI and (b) relative MQI (MQI divided by $\langle \hat{n} \rangle$)
as functions of the degree of squeezing $s$ for the case of
$r_1^2 =$ 0.001 (blue solid), 0.01 (red dashed), and 0.1 (green dot-dashed), respectively.
}
\label{MQI}
\end{figure}

Finally, we consider another kind of size criterion to assess the quality of quantum superposition,
which was very recently proposed by Lee and Jeong \cite{MQI}.
Their measure provides a well-defined criterion for superposition states
by quantifying \emph{the potential of quantum interference} of a state, which can be visualized in phase space.
For this reason, hereafter, we call this measure the ``measure of quantum interference'' (MQI) and define it for a single mode case as
\begin{equation}
{\rm MQI} =
\frac{\pi}{2}  \int \!\! d^2 \beta \; W \!\left(\beta \right)
\! \left[ -\frac{\partial^2 }{\partial \beta \partial \beta ^*} - 1 \right] W\!
\left(\beta \right),
\end{equation}
in terms of Wigner function $W\!\left(\beta \right)$.
A phase-space formalism using a quasi-probability-distribution addresses a general quantum state, so our measure can be used not only for a pure superposition state but also for a mixed state.
Further,
MQI quantitatively manifests unique properties of superposition states.
It gives maximum values for \emph{pure} superposition state and nearly zero values for fully mixed states.
It can be proved that the MQI of a state is bounded by its mean particle number $\langle n \rangle$ and
reaches this maximum value only if it is a pure superposition state.
Consequently, MQI provides useful and sensible information on a superposition state
in regard to the \emph{size of its coherence}.
For the present scheme,
we attempt to assess the quality of generated states by evaluating
the ratio of MQI to the mean photon number $\langle n \rangle$.
This quantity may provide an insight to the  \emph{relative} size of ``superpositionness'' of a quantum state.
It would give a value close to unity if the probed state is close to a pure superposition state and
decrease as the state loses its coherence and hence becomes more mixed.

Figure \ref{MQI} shows the original and the relative values of MQI for the states considered in Fig. \ref{MeanNumber}.
As can be seen from these plots,
the value of MQI increases with the squeezing $s$ for some cases.
However, this is mainly due to the increased size of the state, not due to the increased coherence.
This is confirmed in the plot of \emph{relative} MQI [see Fig. \ref{MQI}(b)],
where one can see that the relative values of MQI monotonically decrease with $s$.
Note that the qualitative behavior of the relative MQI in this plot is very similar to the fidelity in Fig. \ref{P&F}.
In this sense,
the relative MQI seems to provide another useful criterion for the quality of a superposition state
as well as its fidelity.
Moreover,
MQI has an advantage over the fidelity such that
it can assess the quality of a generated state without reference to a target state.

\section{Practical application}
\label{inefficiency}

\begin{figure}[t]
\centerline{ \bf{(a)} }
\vskip0.1cm
\centerline{\includegraphics[width=\columnwidth]{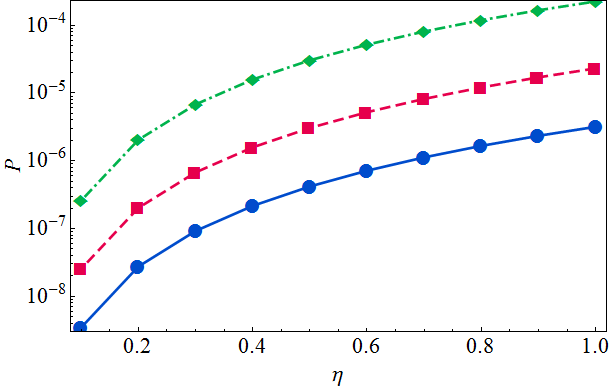}}
\vskip0.1cm
\centerline{ \bf{(b)} }
\vskip0.1cm
\centerline{\includegraphics[width=\columnwidth]{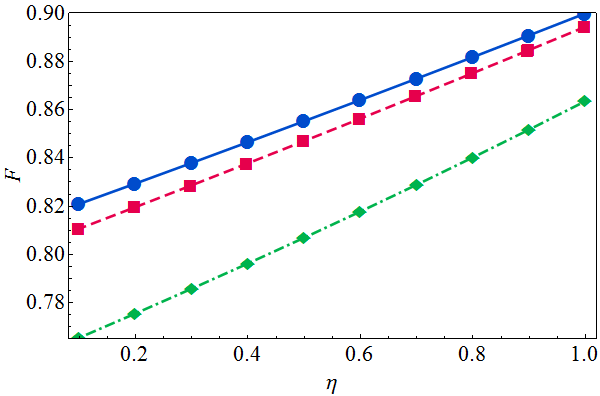}}
\caption{(Color online)
(a) The success probability $P$ and (b) fidelity $F$ as functions of detector efficiency $\eta$ at squeezing $s = 0.16$.
The three curves denote the cases of $r_1^2 =$ 0.001 (blue solid), 0.01 (red dashed), and 0.1 (green dot-dashed) respectively as before.
Notice that as $\eta$ decreases,
$P$ and $F$ also decrease but that,
nevertheless,
the fidelity is relatively robust against the inefficiency compared with the case of probability.
}
\label{P&FIneff}
\end{figure}
\begin{figure}[t]
\centerline{\includegraphics[width=\columnwidth]{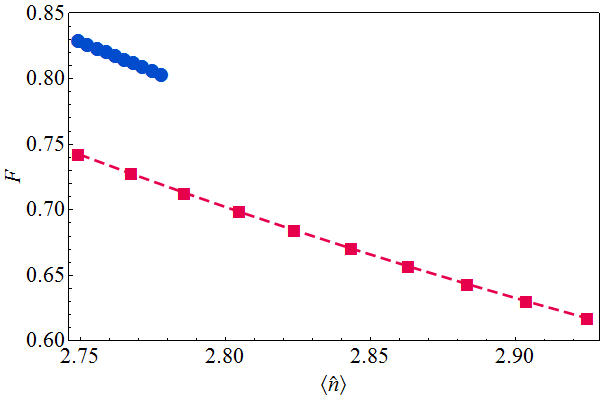}}
\caption{(Color online)
Fidelity $F$ and mean photon number $\langle n \rangle$ for the cases of
$r_1^2 =$ 0.001 (blue solid), 0.01 (red dashed) at $\eta= 1, 0.9, ...,0.1$.
Refer to Fig. \ref{MeanNumber}(b) for the starting values of $F$ and $\langle n \rangle$ at $\eta=1$.
As expected, $F$ decreases with $\eta$ decreasing, while $\langle n \rangle$ slightly increases.
However, the degradation of the fidelity is not very significant in both of the cases $r_1^2 =$ 0.001 and 0.01.
}
\label{n&FIneff}
\end{figure}
In this section, we further investigate the experimental feasibility of our scheme by considering the inefficiency of devices 
and then comparing its performance with the one in Ref. \cite{Gerrits}.
First, we take into consideration the inefficiency of detectors, which can be
modeled by placing an imaginary beamsplitter in front of each detector and varying its transmittance, say $\eta$.
For a perfect (totally inefficient) detector, $\eta=1 ~(0).$
For a low success probability, dark counts may also play another detrimental role.
However, 
it is reported that a coincidence scheme---e.g., which records only the synchronized outcomes of laser pulse signal and a click of a detector---significantly reduces dark count events \cite{Tipsmark}.
There has also been a report of recent progress on on-off type detectors with high efficiency and negligible dark count rate \cite{DarkCount}.

\begin{figure}[t]
\centerline{ \bf{(a)} \hskip2.3cm \bf{(b)} \hskip2.3cm \bf{(c)}}
\centerline{\includegraphics[width=\columnwidth]{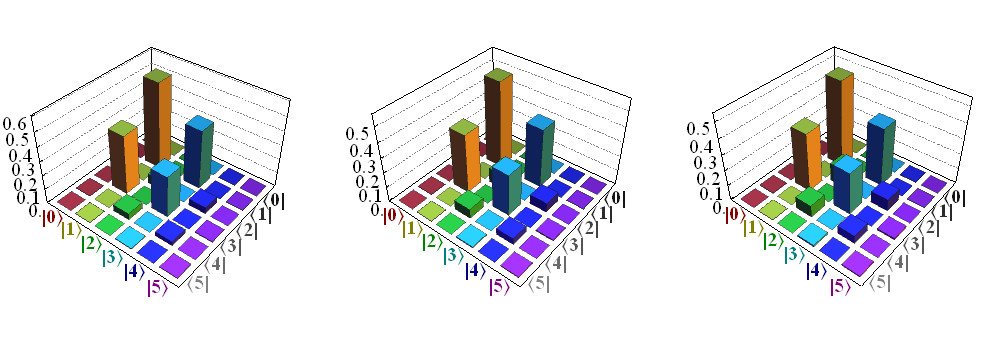}}
\caption{(Color online)
The moduli of the entries of the density matrix of the output state which is generated with
the parameters $s=0.16$,  $r_1^2 = 0.001$, and
the detector efficiency  (a) $\eta=1$, (b) $\eta=0.5$, (c) $\eta=0.1$.
}
\label{DensityMatrix}
\end{figure}

We again calculate the success probability $P$ and the fidelity $F$, but now as functions of efficiency $\eta$ for all the detectors used in our scheme.
We plot $P$ and $F$ for the case of moderate squeezing in Fig. \ref{P&FIneff}.
At $s=0.16$, the fidelity for $r_1^2 = 0.001$ ($r_1^2 = 0.01$) is approximately 90\% (89\%).
Generally, $P$ and $F$ decrease as the detector becomes less efficient, i.e., with $\eta$ decreasing.
However, despite the rapid drop of the probability,
the fidelity is very robust against the detector efficiency and still remains high.
Moreover,
this robustness of the fidelity can also be demonstrated together with the mean photon number.
With the experimental report of Ref. \cite{Gerrits} in mind,
we investigate the effect of detector efficiency $\eta$ for the case of $\langle n \rangle = 2.75$ in \cite{Gerrits} and the fidelity $F = 83\%$ (74\%)
at $r_1^2 = 0.001$ (0.01) in Fig. \ref{MeanNumber}(b).
Figure \ref{n&FIneff} shows both the fidelity and the mean photon number as $\eta$ decreases from 1 to 0.1 by a step of 0.1.
As can be seen from the plot, $F$ is degraded with a decreasing $\eta$ while
$\langle n \rangle$ rather increases in the opposite way.
However, the degradation of $F$ is not very significant in both the cases of $r_1^2 = 0.001$ and 0.01
and the fidelities are still high compared to the one in \cite{Gerrits}.
Even when $\eta$ is lowered to as small as 0.1,
the fidelities are still higher than 59\% reported in \cite{Gerrits},
with mean value slightly higher than $\langle n \rangle = 2.75$.
Further, if we only focus on maximizing the mean photon number,
our scheme can generate a $s$SCS as large as $\langle n \rangle = 3.24~(3.18)$ 
with predetermined squeezing $s=0.2$ and fidelity $F=59\%$ for $r_1^2 =$ 0.001 (0.01) 
even though the detectors have the detector efficiency as low as $\eta = 0.1$.
Finally, 
we also present in Fig. \ref{DensityMatrix} the moduli of the density-matrix entries of the generated state with the squeezing $s=0.16$ and $r_1^2 = 0.001$.
Figure \ref{DensityMatrix} shows that the single- and triple-photon components dominate the other ones and that
even-number components do not appear even when the detector efficiency is very low.
This means that our scheme is reasonably robust against the detector inefficiency and can generate definite odd-parity states even in the presence of imperfect detectors.
These aspects thus illustrate the usefulness of our scheme for generating a viable $s$SCS if implemented in a real experimental setup.

\section{Concluding Remarks}
\label{remark}

We have studied an experimental scheme to generate and grow
a large-size Schr\"{o}dinger-cat-type state using an interferometric setting that implements a coherent superposition of the operations
$\hat{a}\hat{a}^\dag$ and $\hat{a}^{\dag2}$ on an arbitrary input state.
The proposed scheme makes use of an NDPA together with two beam splitters, and the detection of single photons at two output modes conditionally
realizes the superposition operation of our interest.
We investigated the performance of our scheme by employing various criteria such as the success probability, the quantum fidelity,
the mean output energy, and the measure of quantum interference.
We have shown that our scheme can generate an output state of high quality in view of all these criteria
under the use of nonideal on-off detectors for heralding conditional events.
Therefore, together with the experimental achievements reported recently \cite{Gerrits, CommRel}, it seems that our proposed technique can be useful for generating a large-size nonclassical superposition state within existing technology.

\section{Appendix: A}
\label{AppendixA}

In this section, 
we give detailed analytical forms to the classes of states defined in the main text.
For example, 
the state $\ket{s \rm SCS_{\pm}}$ is represented by its wave-function as
\begin{eqnarray}
\langle x \ket{s \rm SCS_{\pm}} &=&
\frac{\mathcal{N}_{\pm}}{\sqrt[4]{\pi \,e^{-2s}}} \Bigg\{ \exp \left[ - \frac{\left( x - \sqrt 2 \alpha e^{-s}  \right)^2 } {2 e^{-2s}} \right] \nonumber \\
&&\pm \exp \left[ - \frac{\left( x + \sqrt 2 \alpha e^{-s}  \right)^2 } {2 e^{-2s}} \right] \Bigg\}
\label{x_sSCS}
\end{eqnarray}
valid up to an overall phase \cite{Barnett}.
Here $\ket{x}$ is the eigenstate of the quadrature operator
$
\hat{x} = \left( \hat{a} + \hat{a}^\dag \right) / \sqrt{2};
$
the detailed form of $\ket{x}$ is given by \cite{Barnett}
\begin{equation}
\ket{x}  = \frac{1}{\sqrt[4]{\pi}} \exp \left[- \frac{1}{2} x^2  + \sqrt{2} x \hat a^\dag  -  \frac{1}{2} \hat a ^{\dag 2} \right] \ket{0},
\label{x_eigenstate}
\end{equation}
where $\ket{0}$ is the vacuum state.
If we set the parameter
$e^{2 s} = 2$, i.e., the case of 3-dB squeezing,
Eq. (\ref{x_sSCS}) is reduced to
\begin{equation}
\langle x \ket{s \rm SCS_{\pm}} =
\mathcal{N}_{\pm} \sqrt[4]{\frac{2}{\pi}} \left[ e^{\left( x - \alpha \right)^2} \pm e^{\left( x + \alpha \right)^2}   \right].
\end{equation}

That is,
\begin{equation}
\langle x \ket{\psi_n} =
\mathcal{N}_{(-)^n} \sqrt[4]{\frac{2}{\pi}} \left[ e^{\left( x - \sqrt{n} \right)^2} + (-1)^n e^{\left( x + \sqrt{n} \right)^2}   \right]
\label{x_psi}
\end{equation}
with
$
\mathcal{N}_{(-)^n} = 1/ \sqrt{2 + 2 (-1)^n e^{-2 n}}.
$

Next,
we can represent $\ket{\phi_n}$ in the number-state basis 
with its wave-function defined in Ref. \cite{Ourjoumtsev07},
\begin{equation}
\langle  x \ket{\phi_n}  =
\sqrt {\frac{2^{2n} n!}{\sqrt \pi  \left( 2n \right)! }} \; x^n e^{ - (1/2) x^2 }.
\label{x_phi}
\end{equation}
We use Eqs. (\ref{x_eigenstate}) and (\ref{x_phi}) to obtain
\begin{eqnarray}
\ket{\phi_n} &=&
\int_{-\infty}^\infty dx \; \ket{x} \langle x \ket{\phi_n} \nonumber \\
&=& (-i)^n \sqrt{\frac{n!}{(2n)!}} ~H_n \! \left[ \frac{i \hat{a}^\dagger}{\sqrt{2}} \right] \ket{0},
\end{eqnarray}
where
$H_n [x]$ is the $n$th-order Hermite polynomial.

\section{Appendix: B}
\label{AppendixB}

We can derive all the detailed expressions of the Wigner functions and the quantum fidelity in Sec. \ref{WignerFormalism}.
First,
a Gaussian distribution of zero mean can be determined solely by its covariance matrix (CM) $V=(V_{ij})$ where $V_{ij} = \frac{1}{2} \langle x_i x_j +  x_j x_i \rangle - \langle x_i \rangle \langle x_j \rangle$ as
\begin{equation}
W_V (x_1,..., x_D) = e^{ -(1/2) {\bf x}^T V^{-1} {\bf x} } / \sqrt{\det (2\pi V)}.
\end{equation}
Here ${\bf x} = (x_1,..., x_D)^T$ is a $D$-dimensional vector,
$T$ denotes the matrix transposition, and $\langle x \rangle$ is the expectation value of $x$.
For example, a single-mode vacuum state can be represented by
\begin{equation}
W_{\rm vac} (\beta) = W_{V_{0}} (\beta_r, \beta_i) ,
\end{equation}
where its CM is given by $V_{0} = \mathcal{I}/4$ with $\mathcal{I}$ being the $2 \times 2$ identity matrix.
Note that a coherent state with amplitude $\alpha$ is given by the same Wigner function with its center just displaced from the origin to $\alpha$ in phase space, i.e.,
$W_{\rm vac} (\beta-\alpha)$.

We now describe the transformation of an input Wigner function to another under the operations such as single-mode squeezing, two-mode squeezing, and beam splitting.
First, a single-mode squeezing transforms the phase-space variables as
\begin{equation}
\beta \to S_1 (s) \beta
\quad \text{where} \quad
S_1 (s) = \left[ \begin{array}{cc} e^{-s} & 0 \\ 0 & e^{s} \end{array} \right].
\end{equation}
Here, $s>0$ ($s<0$) indicates squeezing along the real (imaginary) axis in phase space.
Accordingly, the covariance matrix $V$ under the squeezing can be described by the transformation
$V \to S_1 (s) V S_1^T (s)$
and the output Wigner function thus reads $W_{S_1 (s) V_0 S_1^T (s)} (\beta)$.

Second, two-mode squeezing can be described similarly to the single-mode squeezing above by a transformation matrix
\begin{eqnarray}
S_2 =
\cosh s (\mathcal{I} \otimes \mathcal{I}) + \sinh s (\mathcal{X} \otimes \mathcal{Z} ),\nonumber \\
\text{with} \quad
\mathcal{X} = \left[ \begin{array}{cc} 0~&1\\1~&0  \end{array} \right] \quad \text{and} \quad
\mathcal{Z} = \left[ \begin{array}{rr} 1&0\\0&-1 \end{array} \right].
\end{eqnarray}
The Wigner function of a two-mode squeezed vacuum state in Eq. (\ref{W_ad}) can thereby be represented by
$W_{ad} (\beta_a, \beta_d) = W_{V_{ad}} (\beta_a, \beta_d)$ with
\begin{equation}
V_{ad} = S_2 (V_0 \oplus V_0) S_2^T .
\end{equation}

Third, a two-mode beamsplitting can be described by the transformation
\begin{equation*}
\left[ \begin{array}{c} \beta_A \\ \beta_B \end{array} \right] \to
B_2 (t) \left[ \begin{array}{c} \beta_A \\ \beta_B \end{array} \right]
\quad \text{where} \quad
B_2 (t) = \left[ \begin{array}{cr} t  & -r  \\ r & t \end{array} \right] \otimes \mathcal{I}.
\end{equation*}
Accordingly, two single-mode Gaussian states with their CMs, $V_A$ and $V_B$, are described under beam splitting as
\begin{equation}
W_{V_A \oplus V_B} (\beta_A, \beta_B)
\to
W_{B_2 (t) (V_A \oplus V_B) B_2^T (t)} (\beta_A, \beta_B).
\label{BeamsplitterTransform}
\end{equation}

Tracing over one output mode after the beamsplitting of two Gaussian states gives a single-mode Gaussian state with a modified CM.
Before presenting this formula, we will give some general formulas for integrating Gaussian functions.
The first one is a \emph{partial} integration of a Gaussian Wigner function that can be represented by
\begin{equation}
\int \! d^2 \beta_B W_{V_{AB}} (\beta_A ,\beta_B ) = W_{V_{A(B)}} (\beta_A),
\label{PartialGaussianIntegration}
\end{equation}
where $V_{A(B)}$ denotes the marginal CM for mode $A$, i.e., by removing block matrices related to mode $B$.
The above result simply tells that the marginal distribution must be determined only by its original local CM.
Note that the above and all the following results are not confined to the two-mode case but $A$ and $B$ modes can refer to any collective modes more than one.

Next we present the \emph{partial} integration of the \emph{product} of two Gaussian functions
\begin{eqnarray}
&&\int \! d^2 \beta_B W_{V} (\beta_A ,\beta_B ) W_{\tilde{V}} ( \beta_B -\alpha_B) \nonumber \\
&&= W_{(V+O_A \oplus\tilde{V})} (\beta_A,\alpha_B) ~,
\label{PartialProduct}
\end{eqnarray}
where $O_A$ denotes null matrix associated to mode $A$ and $V$ ($\tilde{V}$) refers to a CM of the two-mode (single-mode) state.
Combining Eqs. (\ref{PartialGaussianIntegration}) and (\ref{PartialProduct}), one can also easily derive another formula,
\begin{equation}
\int \! d^2 \beta~ W_{V} (\beta) W_{\tilde{V}} ( \beta -\alpha)
= W_{V+\tilde{V}} (\alpha) .
\label{ProductGaussianIntegration}
\end{equation}
Now we are ready to obtain a Wigner-function formula associated with a beam splitter.
Integrating Eq. (\ref{BeamsplitterTransform}) over variable $\beta_B$ using the formula  (\ref{PartialGaussianIntegration}), one can get
\begin{eqnarray}
&&\int \! d^2 \beta_B W_{V_A} (t\beta_A + r\beta_B) W_{V_B} (t\beta_B - r\beta_A) \nonumber \\
&=& W_{[B_2 (t) [V_A \oplus V_B) B_2^T (t)]_{A(B)}} (\beta_A)  \nonumber \\
&=& W_{t^2 V_A + r^2 V_B} (\beta_A ).
\end{eqnarray}
If the above two Gaussian states have centers $\alpha_A$ and $\alpha_B$ other than the origin,
the centers of the transformed Gaussian state becomes
$[B_2(t) (\alpha_A, \alpha_B)^T]_{A(B)}$.
In the case of two multi-mode input Gaussian states, e.g., two two-mode states,
the whole process of tracing out after beam splitting can be similarly represented by the transformation
\begin{equation}
W_{V_{AB} \oplus V_{CD}} (\beta_A ,\beta_B, \beta_C,\beta_D)
\to W_{V_{\rm out}} (\beta_A , \beta_B).
\end{equation}
Here, the integration is performed over $\beta_C$ and $\beta_D$ and
\begin{eqnarray}
V_{\rm out}  &=& [B_4 (V_{AB} \oplus V_{CD}) B_4^T ]_{AB(CD)} \nonumber\\
&=& T V_{AB} T^T + R V_{CD} R^T, \nonumber\\
T &=& t_1 \mathcal{I} \oplus t_2 \mathcal{I},~
R= r_1 \mathcal{I} \oplus r_2 \mathcal{I},~
B_4 = \left[ \begin{array}{rr} T & -R \\ R & T \end{array} \right] \nonumber .
\end{eqnarray}
Note again that if there are nonzero centers $(\alpha_A,\alpha_B,\alpha_C,\alpha_D)^T$ in the original Wigner function, those transform to $[B_4 (\alpha_A,\alpha_B,\alpha_C,\alpha_D)^T]_{AB(CD)}$.

Using the general results for Gaussian integration so far,
analytical expressions for all the Wigner functions in the main text are readily obtained.
First, the Wigner functions in Eqs. (\ref{W_a}) and (\ref{P_d}) read
\begin{eqnarray}
W_{a} (\beta_a) &=& \frac{1}{P_d} \Big[ \int \! d^2 \beta_d W_{V_{ad}} (\beta_a ,\beta_d ) \nonumber \\
& & - \pi \int \! d^2 \beta_d W_{V_{ad}} (\beta_a, \beta_d ) W_{V_0} (\beta_d ) \Big] \nonumber\\
&=& \frac{1}{P_d} \left[W_{\cosh 2s V_0} (\beta_a) - \mathrm{sech}^2 s W_{V_0} (\beta_a) \right], \\
P_d &=& 1-  \mathrm{sech}^2 s = \tanh^2 s .
\end{eqnarray}
Next, for the state in Eq. (\ref{Wac}), we obtain
\begin{eqnarray*}
W_{ac} (\beta_a,\beta_c) &=&
\frac{1}{P_d} \Big[W_{S_2 [(\cosh 2s V_0)\oplus V_0]S_2^T} (\beta_a,\beta_c) \nonumber \\
&& - \mathrm{sech}^2 s W_{S_2 [V_0\oplus V_0]S_2^T} (\beta_a,\beta_c) \Big] \nonumber \\
&=&\frac{1}{P_d} \big[W_{V_{ac}} (\beta_a,\beta_c)
- \mathrm{sech}^2 \!s W_{V_{ad}} (\beta_a,\beta_c) \big],
\end{eqnarray*}
with
$V_{ac} =S_2 [(\cosh 2s V_0)\oplus V_0]S_2^T$.
Beam splitting these states gives the Wigner functions in Eqs. (\ref{1stBS}) and (\ref{2ndBS}),
\begin{eqnarray*}
W_{abc} (\beta_a,\beta_b,\beta_c) &=&
\frac{1}{P_d} \big[W_{V_{abc}^{(1)}} (\beta_a,\beta_b,\beta_c) \\
&& - \mathrm{sech}^2 \!s W_{V_{abc}^{(2)}} (\beta_a,\beta_b,\beta_c) \big],\\
W_{abc}' (\beta_a,\beta_b,\beta_c) &=&
\frac{1}{P_d} \big[W_{V_{abc}'^{(1)}} (\beta_a,\beta_b,\beta_c) \\
&& - \mathrm{sech}^2 \!s W_{V_{abc}'^{(2)}} (\beta_a,\beta_b,\beta_c) \big],
\end{eqnarray*}
where
\begin{eqnarray*}
V_{abc}^{(1)} &=& B_{ab} [V_{ac} \oplus (V_{0})_b] B_{ab}^T, \quad
V_{abc}^{(2)} = B_{ab} V_{ad} B_{ab}^T, \\
V_{abc}'^{(1)} &=& B_{bc} V_{abc}^{(1)} B_{bc}^T, \quad
V_{abc}'^{(2)} = B_{bc} V_{abc}^{(2)} B_{bc}^T, \\
B_{ab} &=&  B_2(t_1) \oplus \mathcal{I} ,
\quad \text{and} \quad
B_{bc} = \mathcal{I} \oplus B_2(t) .
\end{eqnarray*}
Finally, on-off detecting on modes $b$ and $c$ gives the output Wigner functions as
\begin{eqnarray*}
W_{\rm out} (\beta_a) &=& \frac{1}{P_d P_{bc}} \sum_{j=1}^{4}
\Big[ A^{(1,j)} W_{V_{\rm out}^{(1,j)}} (\beta_a) \\
&&- \mathrm{sech}^2 \!s~ A^{(2,j)} W_{V_{\rm out}^{(2,j)}} (\beta_a) \Big], \\
P_{bc} &=& \frac{1}{P_d} \sum_{j=1}^{4}
\Big[ A^{(1,j)} - \mathrm{sech}^2\!s~ A^{(2,j)} \Big],
\end{eqnarray*}
where $i=1,2$ and
\begin{eqnarray*}
V_{\rm out}^{(i,1)} &=& V_{a(bc)}'^{(i)}, \\
V_{\rm out}^{(i,2)} &=& [V_{ab(c)}'^{(i)} + O_a \oplus (V_{0})_b]_b^\mathcal{C} , \\
V_{\rm out}^{(i,3)} &=& [V_{a(b)c}'^{(i)} + O_a \oplus (V_{0})_c]_c^\mathcal{C} , \\
V_{\rm out}^{(i,4)} &=& [V_{abc}'^{(i)} + O_a \oplus (V_{0})_b \oplus (V_{0})_c]_{bc}^\mathcal{C}, \\
A^{(i,1)} &=& 1, \\
A^{(i,2)} &=& - \left[2\sqrt{\det [V_{(a)b(c)}'^{(i)} + V_0]}\right]^{-1} , \\
A^{(i,3)} &=& - \left[2\sqrt{\det [V_{(ab)c}'^{(i)} + V_0]}\right]^{-1} , \\
A^{(i,4)} &=& \left[4\sqrt{\det [V_{(a)bc}'^{(i)} + V_0\oplus V_{0}]}\right]^{-1} .
\end{eqnarray*}
Here, the symbol $\mathcal{C}$ denotes \emph{Schur complement} \cite{MA} with the notation 
\begin{eqnarray}
(V_{AB})_A^\mathcal{C} &\equiv& V_{B,B} - V_{B,A} V_{A,A}^{-\!1} V_{A,B}, \nonumber \\
(V_{AB})_B^\mathcal{C} &\equiv& V_{A,A} - V_{A,B} V_{B,B}^{-\!1} V_{B,A},
\end{eqnarray} 
where $V_{A,A}$, $V_{A,B}$ and so on are block matrices in the whole matrix $V_{AB}$ associated to the corresponding modes.

The fidelity in Eq. (\ref{fidelity}) can also be calculated straightforwardly
as the Wigner function of the target $s$SCS in Eqs. (\ref{WSCS}) and (\ref{WsSCS}) can be represented by Gaussian functions, i.e.,
\begin{eqnarray}
W_{\rm SCS} (\beta) &=&
{\cal N}^{\,2}_{-} \big[
W_{V_0} (\beta_r - \alpha, \beta_i) + W_{V_0} (\beta_r + \alpha, \beta_i) \nonumber\\
&& -2 W_{V_0} (\beta) \cos (4\alpha \beta_i) \big], \\
W_{s \rm SCS} (\beta) &=& [W_{\rm SCS} (\beta)]_{V_0 \to S_1 (s') V_0 S_1^T (s')}^{\alpha \to \alpha e^{-s'}} .
\end{eqnarray}
Using these expressions and the formula in Eq. (\ref{ProductGaussianIntegration}),
we obtain the fidelity
\begin{eqnarray*}
F &=& \frac{2\pi{\cal N}^{\,2}_{-}}{P_d P_{bc}} \sum_{j=1}^{4}
\Big[ A^{(1,j)} W_{[V_{\rm out}^{(1,j)}+S_1 (s') V_0 S_1^T (s')]} (\alpha e^{-s'},0) \\
&&- \mathrm{sech}^2 \!s~ A^{(2,j)}W_{[V_{\rm out}^{(2,j)}+S_1 (s') V_0 S_1^T (s')]} (0,i\alpha e^{-s'}) \Big] .
\end{eqnarray*}

\acknowledgments
We thank M. S. Kim for the refined idea of feasible experimental setup,
M. Paternostro for helpful suggestions,
H.-J. Kim for discussion about Gaussian integration, 
and U. L. Andersen and C. R. M\"uller for providing useful information on photodetectors.
H. Nha was supported by NPRP Grant No. 4-554-1-084 from Qatar National Research Fund.
C.-W.L., J.L., and H.J. were supported by the World Class University (WCU) program and
the KOSEF grant funded by the Korea government(MEST) (Grant No. R11-2008-095-01000-0).



\begin{thebibliography}{99}
\bibitem{mistake}
In Ref. \cite{Ourjoumtsev07},
there seems to be a typo in representing $\ket{\phi_2}$ as 
$\ket{\phi_2} = \sqrt{1/3} \ket{0} - \sqrt{2/3} \ket{2}$
[cf. Eq. (\ref{phi2})].
The term of $(-i)^n$ might have been omitted in the momentum representation of $\ket{\phi_n}$:
\begin{equation}
\langle p \ket{\phi_n} =
(-i)^n \sqrt {\frac{2^n n!}{\sqrt \pi (2n)!}} ~H_n \! \left[ \frac{p}{\sqrt 2} \right] e^{-(1/2) p^2}.
\end{equation}
\end{thebibliography}
\end{document}